\documentclass[12pt]{iopart}

\usepackage{iopams}  
\usepackage{graphicx}
\usepackage{epstopdf}
\eqnobysec

\begin{document}

\title[Initial State: Theory Status]{Initial State: Theory Status}

\author{Javier L. Albacete}

\address{$^1$IPNO, Universit\'e Paris-Sud 11, CNRS/IN2P3, 91406 Orsay, France. \\ 
$^{2}$Institut de Physique Th\'eorique,  CEA/Saclay,  91191 Gif-sur-Yvette cedex, France. \\
$^3$Departamento de F\'isica de Part\'iculas, Universidade de Santiago de Compostela. E-15706 Santiago de Compostela, Spain.}
\ead{albacete@ipno.in2p3.fr}
\begin{abstract}
I present a brief discussion of the different approaches to the study initial state effects in heavy ion collisions in view of the recent results from Pb+Pb and p+p collisions at the LHC. 

\end{abstract}

\maketitle

\section{Introduction}

The studies of initial state effects in heavy ion collisions aim at providing a full dynamical description of the colliding system at early times, before the eventual thermalization of the system and the formation of  a Quark Gluon Plasma (QGP).  The importance and practical implications of these studies is manifold. On the soft sector, comprising particle production with small transverse momentum $p_t\le 1\div2$ GeV, they determine the initial conditions (energy and entropy density, initial spatial anisotropy etc ) for the subsequent hydrodynamical evolution of the system. On the hard probes sector, a proper distinction of initial state effects from those originating from the presence of a QGP, or final
state effects, is of vital importance for a proper characterization of the matter produced in heavy
ion collisions, as they may lead sometimes to qualitatively similar phenomena in observables of
interest. Last but not least, another primary goal of initial state studies is to provide a proof that (local) thermalization of the system actually happens over the time scales estimated from hydrodynamical simulations, $\tau_{therm}\sim 0.5\div 1$ fm/c. How such ultra-fast isotropization of the system happens is, arguably, one of the most fundamental open problems in the field of heavy ions.  Our present lack of precise understanding of the pre-equilibrium phase, $0<\tau<\tau_{therm}$, motivates the bypass of this part of the dynamics in most phenomenological works for bulk particle production.
  
Below i briefly review the theoretical interpretation of the new data on Pb+Pb collisions at the LHC, with emphasis in the Color Glass Condensate approach. 

\section{Coherence effects}
\begin{figure}
\hskip 4.5cm
\includegraphics[width=70mm]{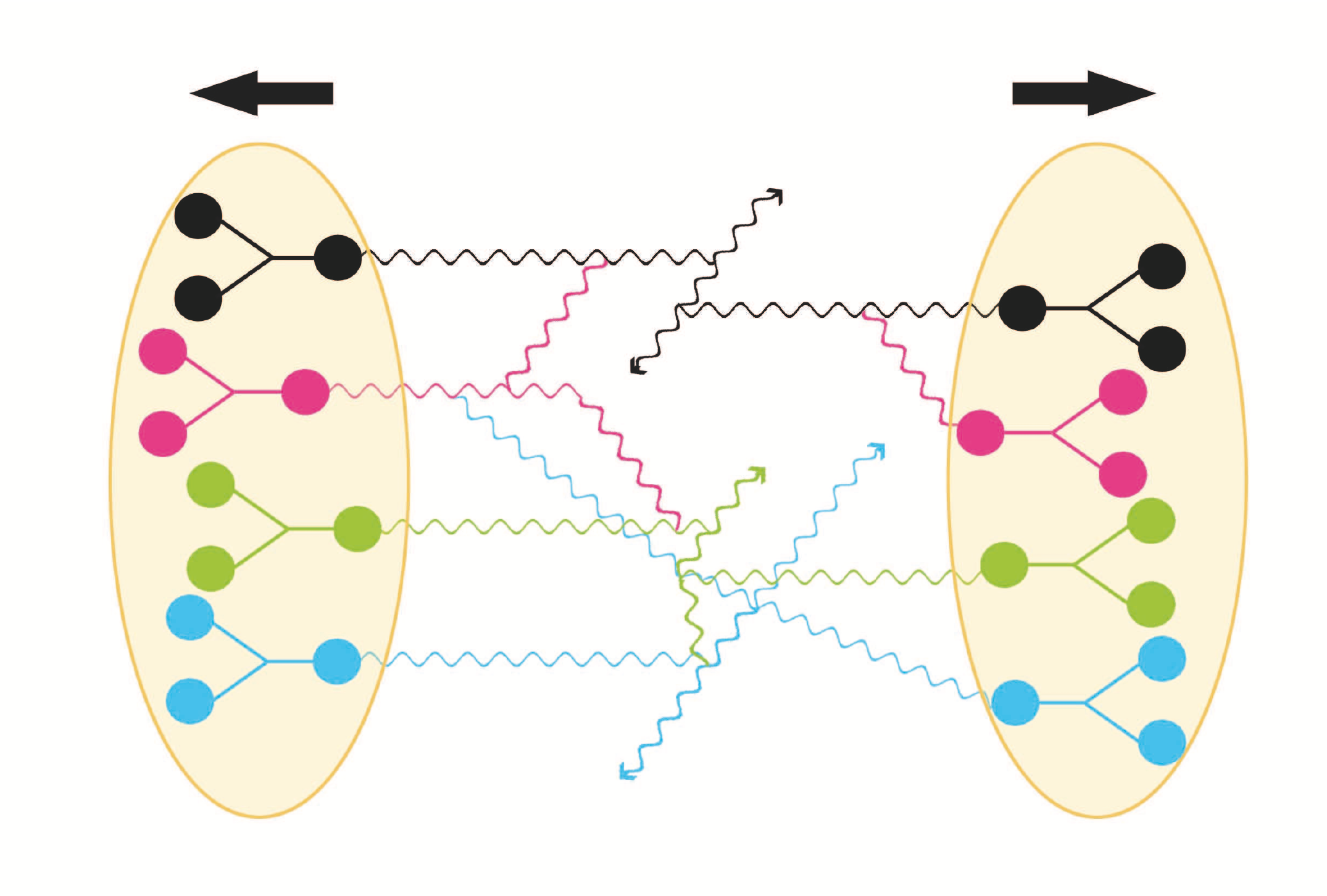}
\caption{Pictorial representation of coherence effects in the early stages of heavy ion collisions.}
\label{Last}
\end{figure}

A main lesson learnt from experimental data collected in Au+Au and Pb+Pb collisions at RHIC and the LHC respectively is that bulk particle production in ion-ion collisions is very different from a simple superposition of nucleon-nucleon collisions. Such is evident in terms of the measured charged particle multiplicities, which exhibit a strong deviation from the scaling with the number of nucleon-nucleon collisions: $\frac{dN^{AA}}{d\eta}(\eta=0)\ll N_{coll} \frac{dN^{AA}}{d\eta}(\eta=0)$. This observation leads to the conclusion that strong coherence effects among the constituent nucleons, or the relevant degrees of freedom at the nucleon level, must be present during the collisions process. Indeed, any of the phenomenological models who successfully describe data include strong coherence effects. While a detailed discussion of the different prescriptions found in the literature to account for coherence effects is beyond the scope of this brief review, one can identify in different models coherence effects at the level of the wave function and also at the level of primary particle production, sketched in Fig (\ref{Last}). To the first category correspond the nuclear shadowing (in a partonic language) or the percolation and string fusion (in non-perturbative approaches). In both cases, when different constituents, whichever the degrees of freedom chosen are, overlap in phase space according to some geometric criterium, recombination of such constituents happen, thus reducing the total number of scattering centers entering the collision process. Similar phase-space arguments motivate the implementation of energy-dependent cut-offs to regulate independent particle production from different sources, normally a working hypothesis in most Monte Carlo event generators for heavy ion collisions.

\subsection{The Color Glass Condensate (CGC) approach}
The CGC attempts at providing a complete, QCD-based, dynamical description of the coherence effects sketched above (for a review see, e.g., \cite{Gelis:2010nm}). On one hand, the gluon {\it shadowing} is taken into account through non-linear renormalization group equations, the BK-JIMWLK equations, that describe the change in hadron structure towards small Bjorken-$x$ (equivalently, higher collision energies). The non-linear terms in the BK-JIMWLK equations reflect the probability of gluon-gluon recombination processes in the high density regime. Those terms ensure unitarity of the theory and tame the growth of gluon densities towards small-$x$. They also imply the emergence of a dynamical transverse momentum scale, the saturation scale $Q_s$, such that gluon modes with transverse momentum $k_t\le Q_s(x)$ are in the saturation regime \cite{Albacete:2007yr}. 

Saturation of gluon densities is equivalent to the presence of strong color fields, parametrically of the order of the inverse of the coupling, $\mathcal{A}(x)\sim 1/g$. Thus, although the insertion of new sources in the diagrams for calculating particle production seem a priori suppressed by powers of the coupling constant, such additional contributions are compensated by the strength of the color fields, i.e. terms of the order $g\, \mathcal{A}(x)\sim \mathcal{O}(1)$ must be resummed to all orders. This implies a rearrangement of the perturbation series in the high-density regime. To first approximation, all leading terms can be resummed by solving the classical Yang-Mills equations of motion (CYM), with the valence degrees of freedom of the two colliding nuclei acting as external non-dynamical (in the eikonal approximation) sources.  However, supplementing the classical methods with information on the quantum evolution of the wave function of the colliding nuclei, necessary to safely extrapolate from RHIC to LHC energies, is presently a difficult task since it requires the numerical resolution of the full JIMWLK equations. Although important steps have been taken in that direction \cite{Lappi:2011ju}, such program is not yet fully realized. 
Rather, the most common approach to describe initial gluon production relies in the use of $k_t$-factorization, strictly valid only for {\it dilute-dense} scattering. There, the differential cross-section for small-$x$ gluon production is given by the convolution of the unintegrated gluon distributions (ugd's) of projectile and target, $\varphi(k_t,x,b)$,

\begin{figure}
\hskip 1cm 
\includegraphics[width=70mm]{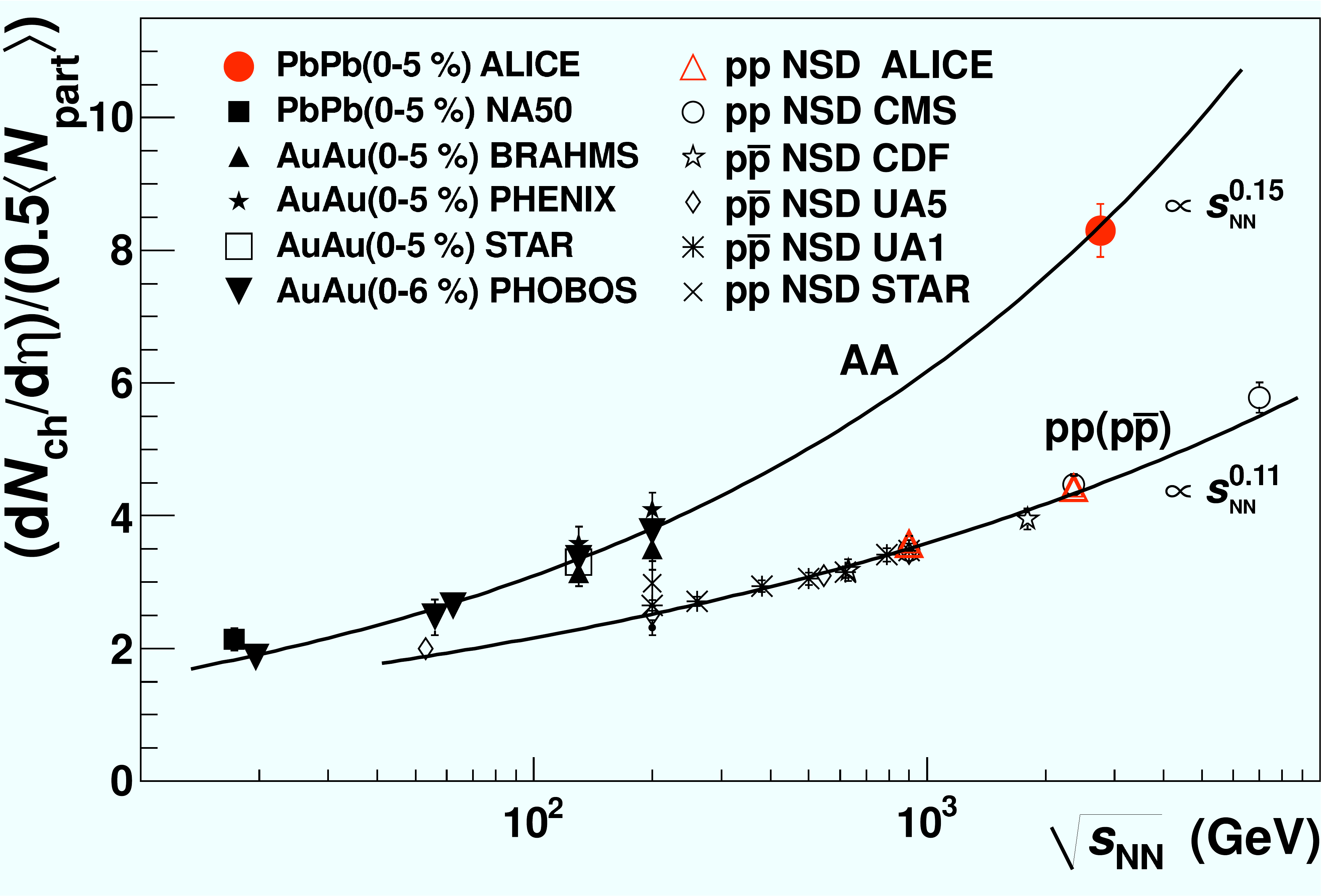}
\hskip 1cm
\includegraphics[width=67mm]{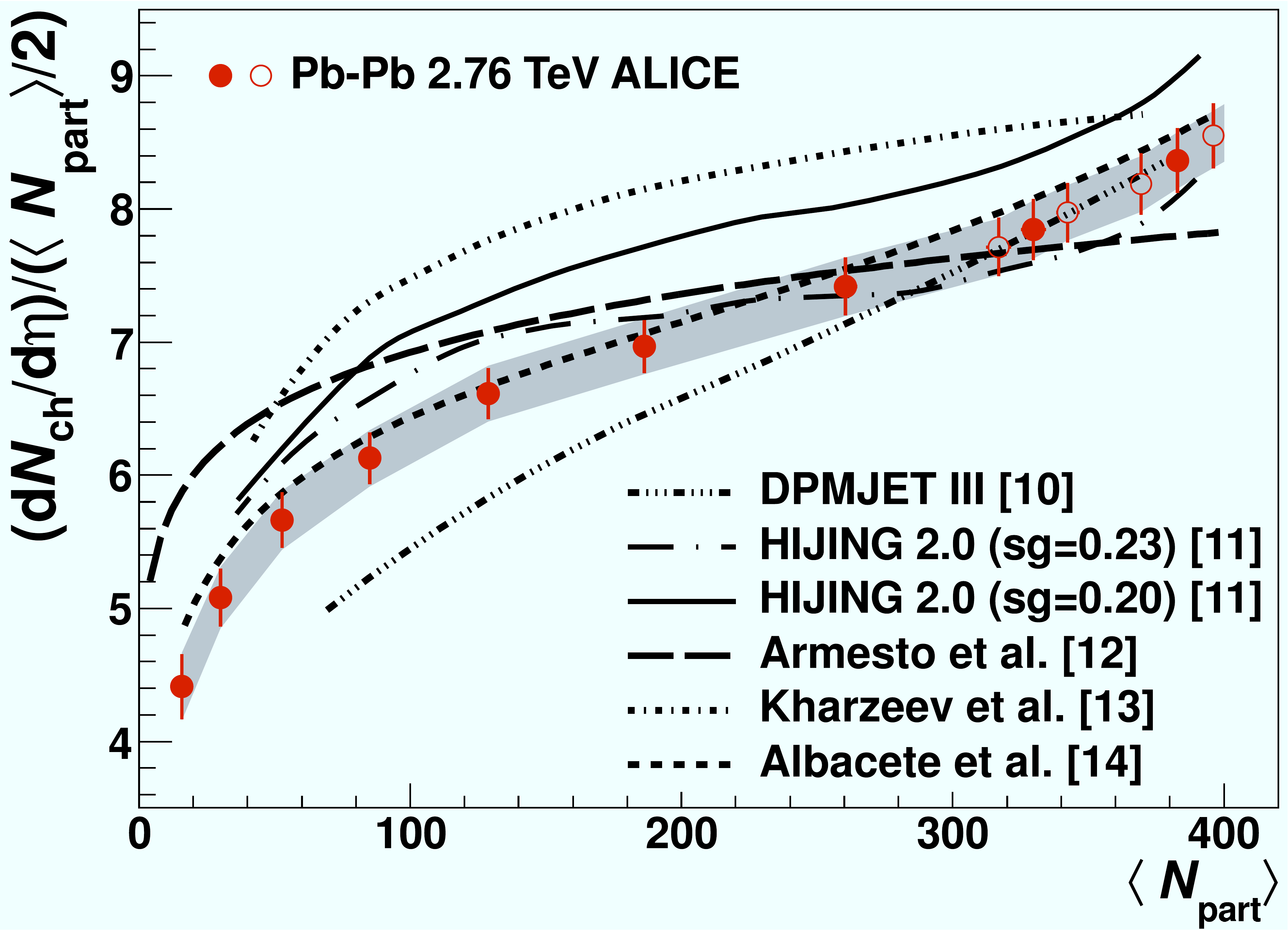}
\caption{Energy (left) and centrality (right) dependence of mid-rapidity multiplicities in A+A and p+p collisions. Figures and LHC data by the ALICE collaboration \cite{Aamodt:2010pb,Aamodt:2010cz}.}
\end{figure}

\begin{equation}
\hskip -2.5cm
\frac{d\sigma^{A+B\to g}}{dy\, d^2p_t\, d^2R} =  \frac{2}{C_F} 
\frac{1}{p_t^2} \int \frac{d^2k_t}{4}\int d^2b\, 
\alpha_s(Q)\,
\varphi(\frac{|p_t+k_t|}{2},x_1;b)\,
\varphi(\frac{|p_t-k_t|}{2},x_2;R-b)~,
\label{eq1}
\end{equation}
where $x_{1(2)}=(p_t/\sqrt{s_{NN}})\exp(\pm y)$, and $p_t$  and $R$ the transverse momentum and impact parameter of the produced gluon. In turn, the $x$-dependence of the ugd's can be calculated by solving the BK equation, which corresponds to the large-$N_c$ of the full JIMWLK and is more amenable to both analytical and numerical analysis. This allows a controlled theory extrapolation to higher energies or more forward rapidities. In this approach, final state hadron multiplicities are taken to be proportional to initially produced minijets (gluons with momenta $p_t\sim Q_s$) through the local parton-hadron duality in the form of a {\it fugacity} factor.  

\section{Bulk features of multiparticle production}
Two main features of the new LHC data \cite{Aamodt:2010pb,Aamodt:2010cz} can be highlighted: First, the energy dependence of mid-rapidity multiplicities in A+A collisions is well reproduced by a power law, $dN^{ch}/d\eta(\eta=0)\sim s^{0.15}$. This observation seems to rule out the logarithmic trend observed for lower energies data and is in generic agreement with pQCD based approaches.  Somewhat unexpectedly, LHC data seem to indicate a stronger energy dependence in mid-rapidity multiplicities in p+p collisions. While no clear explanation of this is observation is yet available, several possibilities have been recently proposed based on the ideas of additional entropy production in the pre-equilibrium phase \cite{Baier:2011ap},  enhanced parton showers in A+A collisions due to the larger average transverse momentum of the initially  produced minijets compared to p+p collisions \cite{Levin:2011hr} or to non-trivial high-$Q^2$ effects intertwined with impact parameter dependence \cite{Lappi:2011gu}.
Second, the centrality dependence of multiplicities is, up to an overall scale factor, very similar to the one observed at RHIC. This suggests a factorization of the energy and centrality dependence of the multiplicities, which, in turn, admits a natural explanation in the CGC formalism. There, mid-rapidity multiplicities rise proportional to the saturation scale of the colliding n uclei which, in a first approximation, is proportional to the local nuclear density or, equivalently, to the number of participants:

\begin{equation}
\frac{dN}{d\eta}(\eta=0) \sim Q_s^2(x,b) \sim \sqrt{s}^{\lambda}\,N_{part}\,.
\label{eq2}
\end{equation}

\begin{figure}

\hskip 1cm 
\includegraphics[width=70mm]{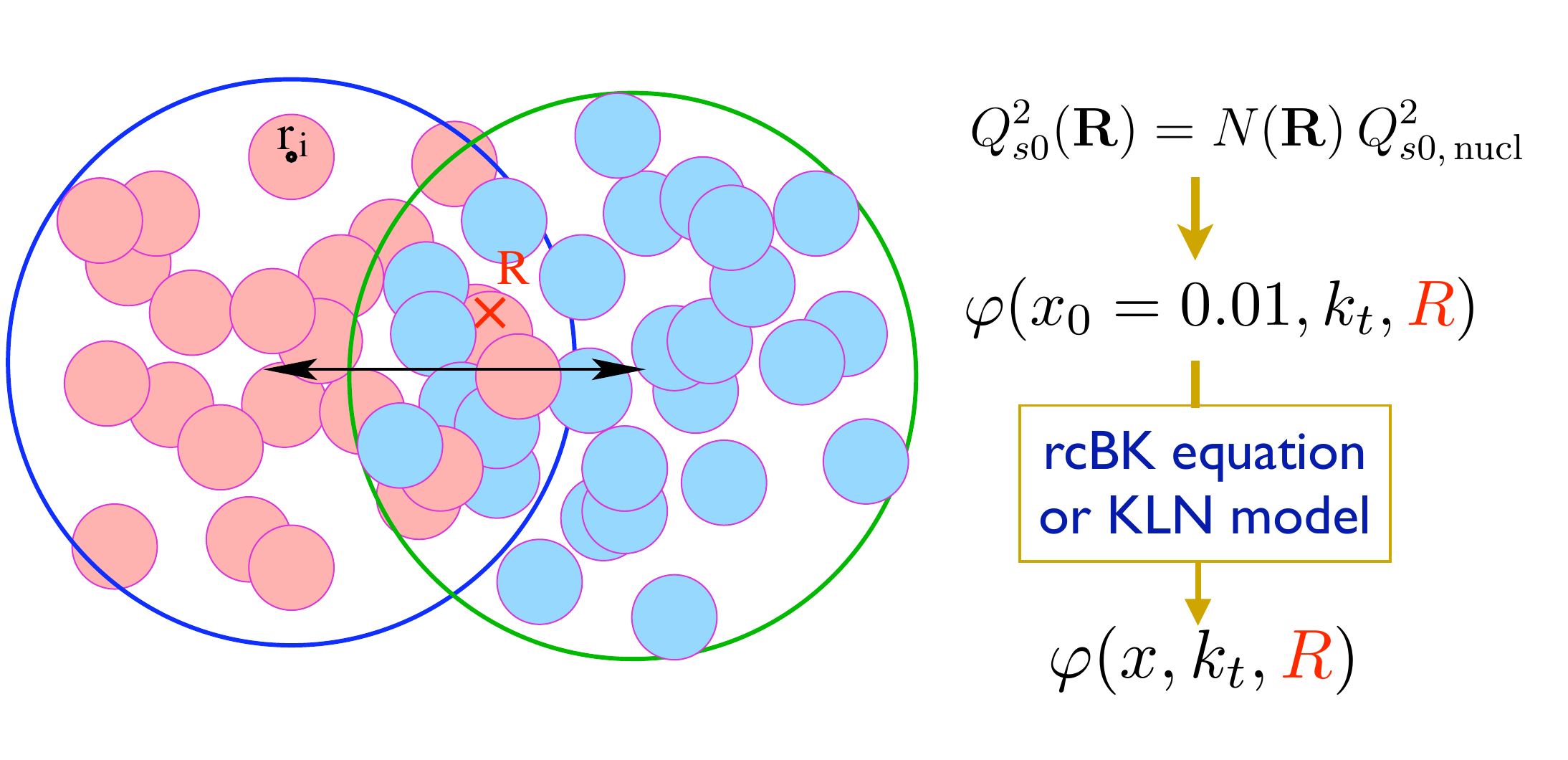}
\includegraphics[width=70mm]{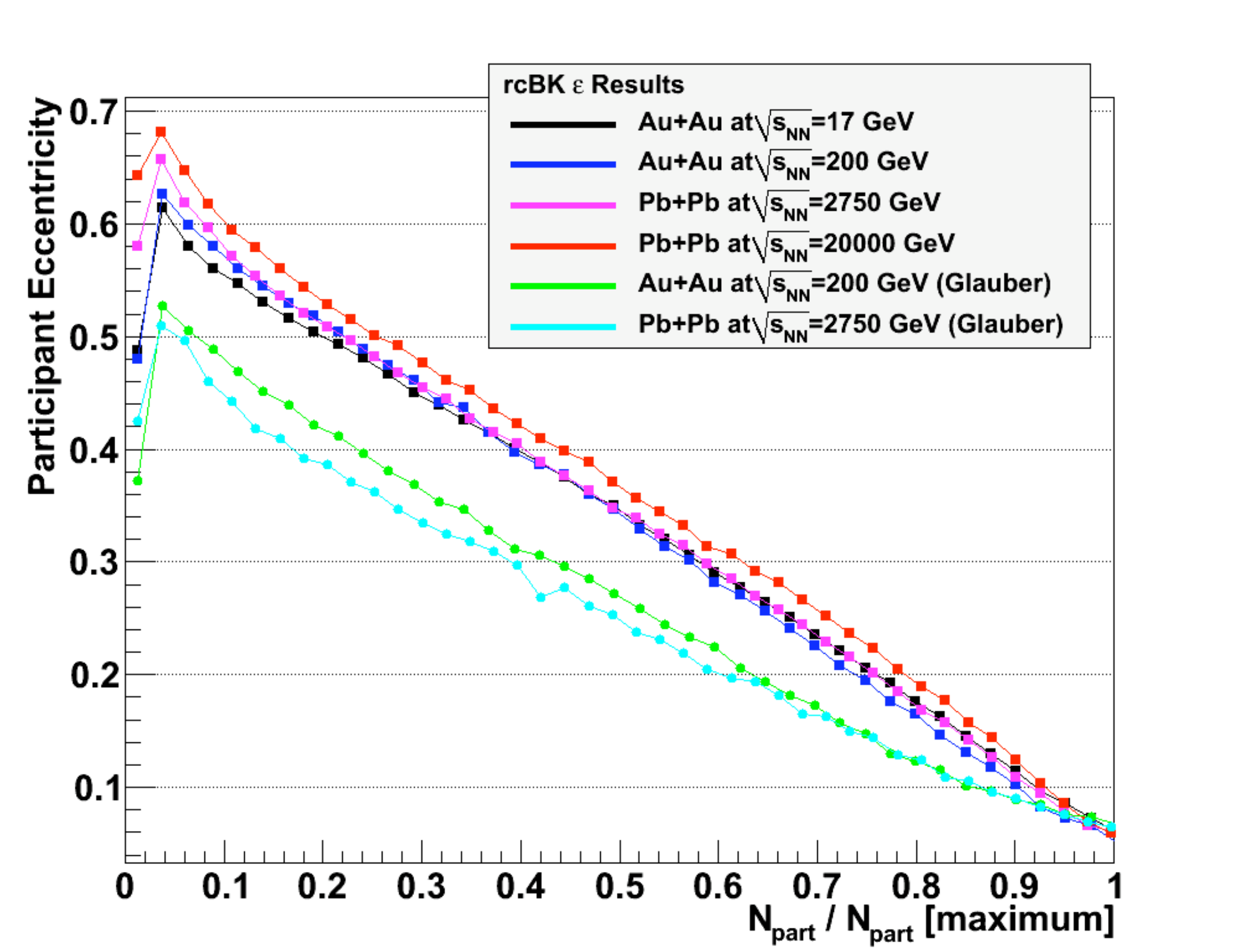}
\hskip 1cm
\label{mc}
\caption{Left: Sketch of the CGC Monte Carlo implementation. Right: Participant eccentricity as a function of centrality for rcBK-CGC and Glauber initial conditions at different collisions energies (courtesy of J. Nagle).}
\end{figure}

Fig. 1 (right) includes comparison of different Monte Carlo \cite{Deng:2010xg,Bopp:2007sa} and saturation based calculations \cite{Kharzeev:2004if,Armesto:2004ud,ALbacete:2010ad} to data. 
A best description of data is given by the MC-rcBK approach of \cite{ALbacete:2010ad} and in the MC HIJING approach \cite{Deng:2010xg}, which includes a very strong impact parameter dependent shadowing. Aside of implementation details, as infrared regularization or the possible origin of scaling violations,  different CGC approaches differ mainly in their input for the ugd's in (\ref{eq1}): analytical models \cite{Kharzeev:2004if}, models constrained from e+p and e+A  data \cite{Armesto:2004ud} or solutions from the running coupling BK equation, a recent and important theoretical development in the CGC framework (see \cite{Albacete:2007yr} and references therein). These choices lead to slightly varying effective values of the exponent $\lambda\sim{0.2\div 0.3}$ in   (\ref{eq2}) and partially explain the spread of {\it saturation} based predictions. Other distinguishing feature among saturation models is the treatment of impact parameter dependence. It varies from describing the nucleus by a single, average saturation scale, to mean field approaches with explicit b-dependence, to Monte Carlo methods. The latter are the best suited to account for the fluctuations in the initial geometry of the collision, a crucial ingredient for the subsequent analysis of multiparticle correlations, flow, etc. Presently there are two variants of CGC Monte Carlo methods, the MC-KLN \cite{Drescher:2007ax} and MC-rcBK \cite{ALbacete:2010ad}, schematically represented in Fig. (\ref{mc}). Their geometric set up is identical in both cases, differing in the dynamical input for the $(x,k_t)$-dependence of the ugd's, either the KLN model\cite{Kharzeev:2004if} or the rcBK approach. Particle production is then calculated according to $k_t$-factorization. A crucial ingredient in both cases is the ugd for a proton at the largest value of $x$ considered (typically $x\sim 0.01$), which in the rcBK approach is constrained by global fits to e+p HERA  data on structure functions \cite{Albacete:2010sy}. 

As shown in Fig. (\ref{mc}),  MC-CGC approaches systematically yield a larger initial eccentricity than Glauber ones. However, it is unclear to what extent such is a robust property of the CGC formalism (see for instance the CYM studies in \cite{Lappi:2006xc}) or an artifact of their current implementation in phenomenological works. Also, it should be kept in mind that initial spatial gradients are very sensitive to particle production in the dilute periphery of the collision area. There, the applicability of the CGC formalism, which relies in the presence of high gluon densities, is not guaranteed. Rather, non-perturbative effects may play an important role in that region. In that sense, aspects of the modeling in the MC tools used, such as the relevant sources of fluctuations or the model for the nucleon geometry themselves (e.g thick discs vs. gaussian) may be more relevant for the determination of initial eccentricities than the choice of underlying description for initial particle production, i.e CGC or Glauber.
Also, studies of higher flow harmonics presented in this conference (by e.g. the PHENIX or ALICE collaborations) do not clearly favor CGC or Glauber approaches to describe the initial state of the collision.

\section{High $p_t$-particle production}

\begin{figure}
\hskip 0cm 
\includegraphics[width=72mm]{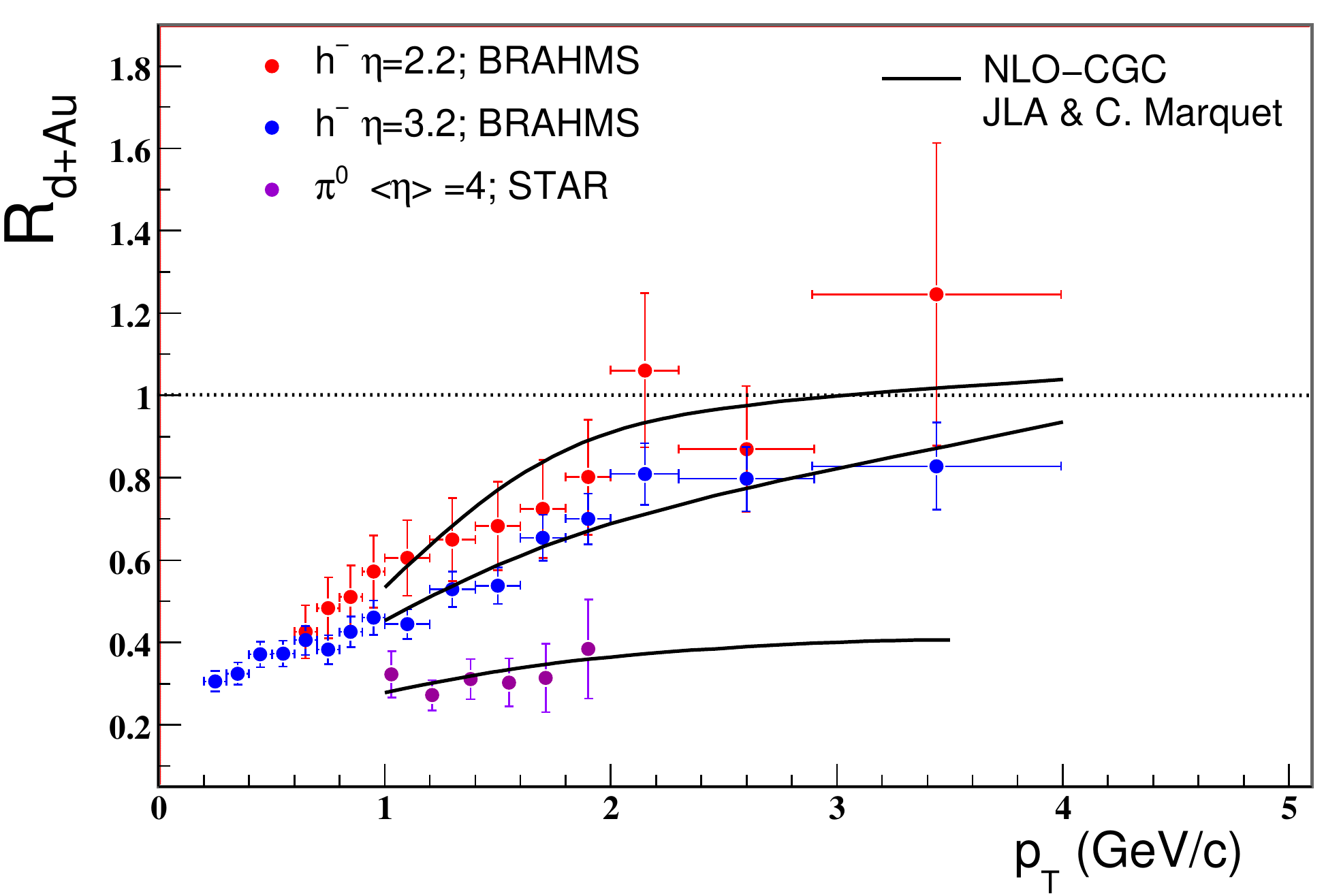}
\hskip 1cm
\includegraphics[width=72mm]{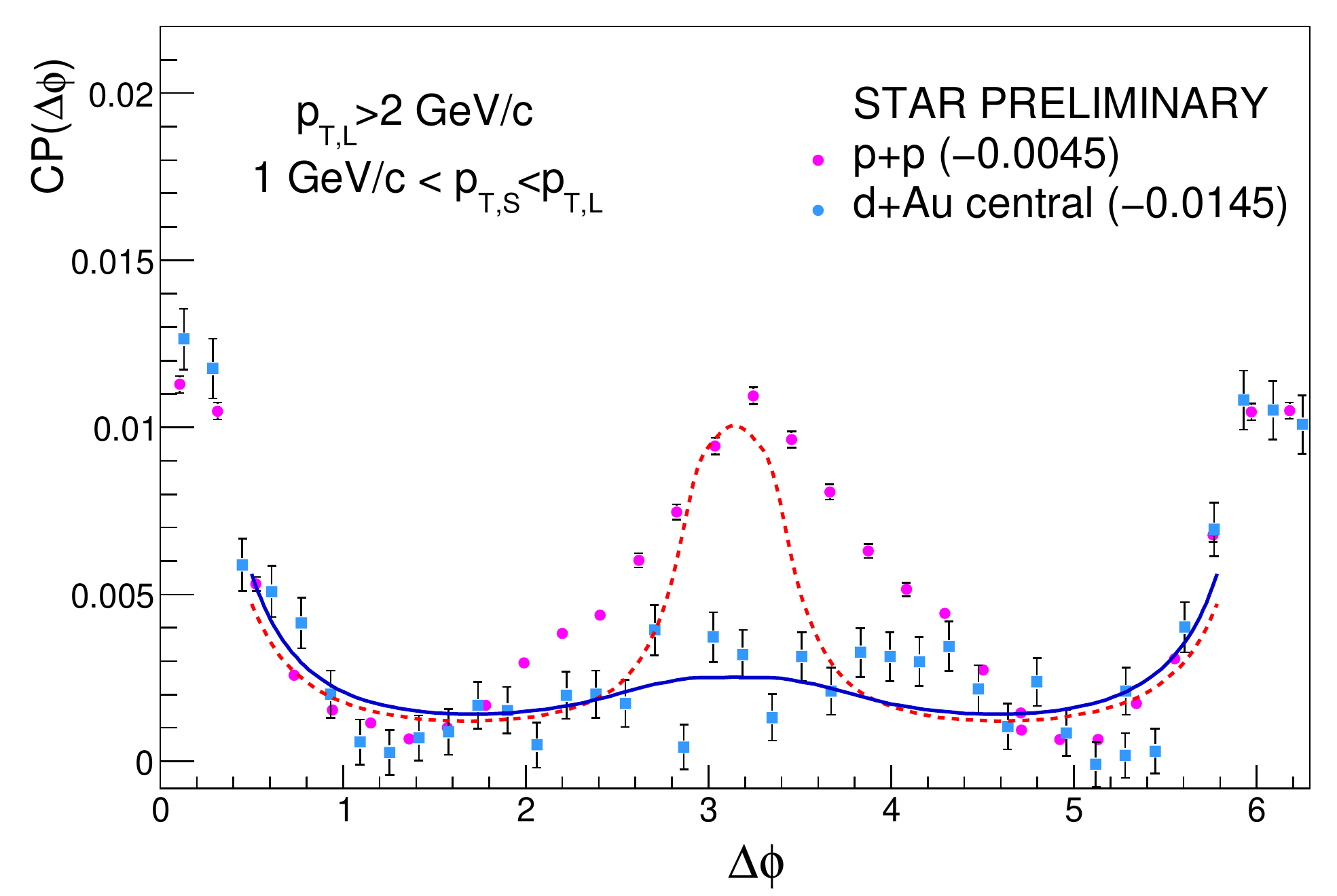}
\label{forward}
\caption{Suppression of forward single inclusive yields (left) and di-hadron azimuthal correlations (right) compared to CGC calculations \cite{Albacete:2010bs,Albacete:2010pg}. Data by BRAHMS and STAR collaborations, respectively  .}
\end{figure}

The observation of suppression phenomena in forward measurements in d+Au collisions at RHIC has been consistently explained in terms of CGC effects \cite{Albacete:2010bs,Albacete:2010pg}. Thus, both the continuous depletion of nuclear modifications factors in single inclusive production (Fig. (\ref{forward}) left) and the disappearance of forward azimuthal correlations (Fig. (\ref{forward}) right) can be interpreted as due to the presence of a dynamical, semi-hard scale in the nucleus wave function. The use of solutions of the running coupling BK equation to describe the $x$-evolution of the nuclear wave function turns out to be necessary in order to attain a good quantitative description of both observables. A caveat in both calculations is that high-$x$ effects, such as energy loss effects or the effects \cite{Kopeliovich:2005ym} of multi-parton interactions, argued to be important in the forward region \cite{Strikman:2010bg}, are not taken into account. However, preliminary measurements presented by the STAR collaboration in this conference on neutron tagged events, thus reproducing {\it true} p+Au collisions, confirm the suppression of azimuthal correlations, thus eliminating the uncertainty on the role of multi-parton interactions and lending support to the CGC interpretation.
These studies have been taken as a baseline to estimate the nuclear modification factors for single inclusive hadron production at the LHC, predicting a similar suppression in LHC p+Pb collisions at mid rapidity to the one observed in forward rapidities at RHIC. This is in contrast with predictions based on the use of collinear factorization and nuclear pdf's (see eg \cite{Salgado:2011wc}). However, the predictions of \cite{Albacete:2010bs} suffer of an uncertainty on the absolute normalization, due to the fact that the nuclear and proton saturation scales were fitted independently to data. The requirement that initial state, saturation effects should disappear at high enough $p_t$ demands that they should be related through the thickness function. Such could be achieved by redoing the phenomenological analyses of \cite{Albacete:2010bs,Albacete:2010pg} in the MC-CGC set up.

\section{Conclusions}
Important steps have been taken over the last years in promoting the CGC framework to a predictive and quantitative phenomenological tool. Such has been possible through the systematic implementation of global fit and Monte Carlo methods and, more importantly, through an intense theoretical work in the determination of higher order corrections to the formalism, including running coupling corrections to non-linear evolution equations and also to particle production processes. 
First available data on features of bulk particle production in Pb+Pb collisions are in good agreement with improved CGC expectations, but they are also compatible with Monte Carlo event generators. Both approaches have in common the inclusion of strong coherence effects. An exhaustive analysis of forthcoming more differential observables is needed to better discriminate between the two.

One can identify two most urgent tasks in order to improve the predictive power of CGC calculations: i) Further inclusion of high-$x$ and high-$Q^2$ effects presently not accounted for in the formalism. This would provide a matching with the collinear DGLAP formalism. ii) Improve our knowledge of the impact parameter dependence of nuclear densities. This is a problem related to the physics of confinement. In that sense a p+Pb run at the LHC would provide extremely valuable empiric information to better calibrate initial state effects not only for hard probe production, but also to further constraint models for bulk particle production.

\ack
I would like to thank the organizers of Quark Matter 2011 for their invitation to such an interesting conference. This work is supported by a Marie Curie Intra-European Fellowship (FP7- PEOPLE-IEF-2008), contract No. 236376 and by by  Ministerio
de Ciencia e Innovaci\'on of Spain (project FPA2008-01177), Xunta de
Galicia (project PGIDIT10PXIB 206017PR),  and by FEDER.


\end{document}